\documentclass[twocolumn,prd,nofootinbib,aps,floats,floatfix,amsmath,amssymb,secnumarabic]{revtex4} %
\usepackage[final]{graphicx}
\usepackage{amsmath}
\usepackage{bbm}
\usepackage{amsfonts}
\usepackage{amssymb}
\usepackage{latexsym}
\usepackage{graphicx}
\usepackage[english]{babel}
\usepackage{multirow}
\usepackage{float}
\usepackage{url}
\usepackage{hyperref}
\usepackage{slashed}
\usepackage{xcolor} 

\def\sfrac#1#2{{\textstyle{#1\over #2}}}
\newcommand{\be}{\begin{equation}}
\newcommand{\ee}{\end{equation}}
\newcommand{\ba}{\begin{array}}
\newcommand{\ea}{\end{array}}
\newcommand{\bea}{\begin{eqnarray}}
\newcommand{\eea}{\end{eqnarray}}

\newcommand{\de}{{\mathbf e} }
\newcommand{\pd}{{\mathbf p} }
\newcommand{\dH}{{\mathbf H} }

\begin{document}

\title{3.5 keV X-rays as the ``21 cm line'' of dark atoms,\\
and a link to light sterile neutrinos}
\author{James M.\ Cline}%\footnote{jcline@physics.mcgill.ca}}
\author{Zuowei Liu}%\footnote{zuoweiliu@physics.mcgill.ca}}
\author{Guy D.\ Moore}%\footnote{zuoweiliu@physics.mcgill.ca}}
\affiliation{Department of Physics, McGill University,
3600 Rue University, Montr\'eal, Qu\'ebec, Canada H3A 2T8}
\author{Yasaman Farzan}
\affiliation{School of physics, Institute for Research in Fundamental 
Sciences (IPM)\\
P.O.Box 19395-5531, Tehran, Iran}
\author{Wei Xue}%\footnote{xueweiphy@gmail.com}}
\affiliation{INFN, Sezione di Trieste, SISSA, via Bonomea 265, 
34136 Trieste, Italy}

\begin{abstract}  
The recently discovered 3.5 keV X-ray line from extragalactic sources 
may be evidence of dark matter scatterings or decays.  We show that
dark atoms can be the source of the emission,  through their hyperfine
transitions, which  would be the analog of 21 cm radiation from a dark
sector.  We identify two families of dark atom models that 
match the X-ray observations and are consistent 
with other constraints.  In the first, the hyperfine excited state 
is long-lived compared to the age of the universe, and the dark atom
mass is relatively unconstrained; dark atoms could be strongly
self-interacting in this case.  In the second, the excited state is
short-lived and viable models are parameterized by
the value of the dark proton-to-electron mass ratio $R$:
for $R = 10^2-10^4$,
the dark atom mass is predicted to be in
the range $350-1300$ GeV, with fine structure constant 
$\alpha'\cong 0.1-0.6$.  In either class of models, the dark photon 
is expected to be massive with $m_{\gamma'} \sim$ 1 MeV and decay into $e^+ e^-$. 
Evidence for the model could come from direct detection of the dark
atoms.   In a natural extension of this framework,  the 
dark photon could decay predominantly into invisible particles,
for example $\sim 0.5$ eV sterile neutrinos,  
explaining the extra radiation degree of freedom recently
suggested by data from BICEP2, while remaining compatible with BBN.

\end{abstract}
\maketitle

%\section{Introduction}  

{\bf\noindent 1.\ Introduction.} Evidence for a 3.5 keV X-ray line has been found in XMM-Newton data
from the Andromeda galaxy and the Perseus galaxy cluster
\cite{Boyarsky:2014jta}, as well as in the stacked spectra of 73
galaxy clusters \cite{Bulbul:2014sua}.   Since a line at this energy
is not attributable to any known atomic transitions, there has been
considerable interest in trying to explain it in terms of dark matter
scattering or decays \cite{Ishida:2014dlp}.  %-\cite{Babu:2014pxa}. 
In most of these models, the dark matter is very light, with mass
of order a few keV.  Alternatively, an excited metastable state of 
dark matter could decay to its ground state with the emission of a
3.5 keV photon, or an unstable state of this type could be excited
through inelastic dark matter scatterings 
\cite{Finkbeiner:2014sja,Frandsen:2014lfa}.

In the present note we point out that dark atoms provide an
interesting example of the latter two kinds, due to the relative ease of
inducing hyperfine excitations through their
self-scatterings. This idea was also considered in 
\cite{Frandsen:2014lfa}, but the difficulties for massless dark photons
that we highlight and overcome here were not spelled out definitively 
in that paper.
Although atomic dark matter is an old idea, originally arising in theories of
mirror symmetry (see \cite{Foot:2004pa} for a review), it has experienced a revival in the larger context
of hidden sectors \cite{Kaplan:2009de}-\cite{Cline:2013zca}, since any U(1) gauge symmetry with sufficiently
light mediators should give rise to atom-like bound states.   We find it
appealing that atomic dark matter can naturally explain the X-ray
line, without giving up on the WIMP paradigm, by identifying the
low 3.5 keV energy scale with the hyperfine splitting:
\vskip-0.5cm
\be
	\Delta E = \frac{8}{3}\alpha'^4 {m_{\de}^2 m_{\pd}^2\over
	m_\dH^3} = 3.5{\rm\ keV}
\label{dE}
\ee
Here $\de$, $\pd$ and $\dH$ respectively denote the dark electron, 
proton (assumed to be elementary particles), and
hydrogen atom, with fine structure constant $\alpha'$ and $m_\dH = 
m_{\de} + m_{\pd}$.   

In the following we identify two regimes in which the decay of the 
hyperfine excited state can explain the X-ray line, both requiring the
dark photon to be massive with $m_{\gamma'}\sim 1$ MeV.
If the decays are relatively fast, 
then  $m_\dH\sim 350-
1300$ GeV and $\alpha'\sim 0.1-0.6$ 
for $R\equiv m_\pd/m_\de=10^2-10^4$, in order to satisfy 
eq.\ (\ref{dE}) plus the
observed strength of the 3.5 keV line, and constraints from perturbativity
and from recombination in the dark sector. 
%(Larger $R$ may
%also be possible, but numerical computations of the cross sections
%have not been carried out for $R>10^4$.)  
If the decays are slower
$\sim 10^{-17}$\,s$^{-1}$, the line intensity depends on the fractional
electric charge of the dark constituents, and there is more
freedom in choosing consistent parameter values.

\begin{figure*}[t]
\centerline{
\includegraphics[width=1.05\columnwidth]{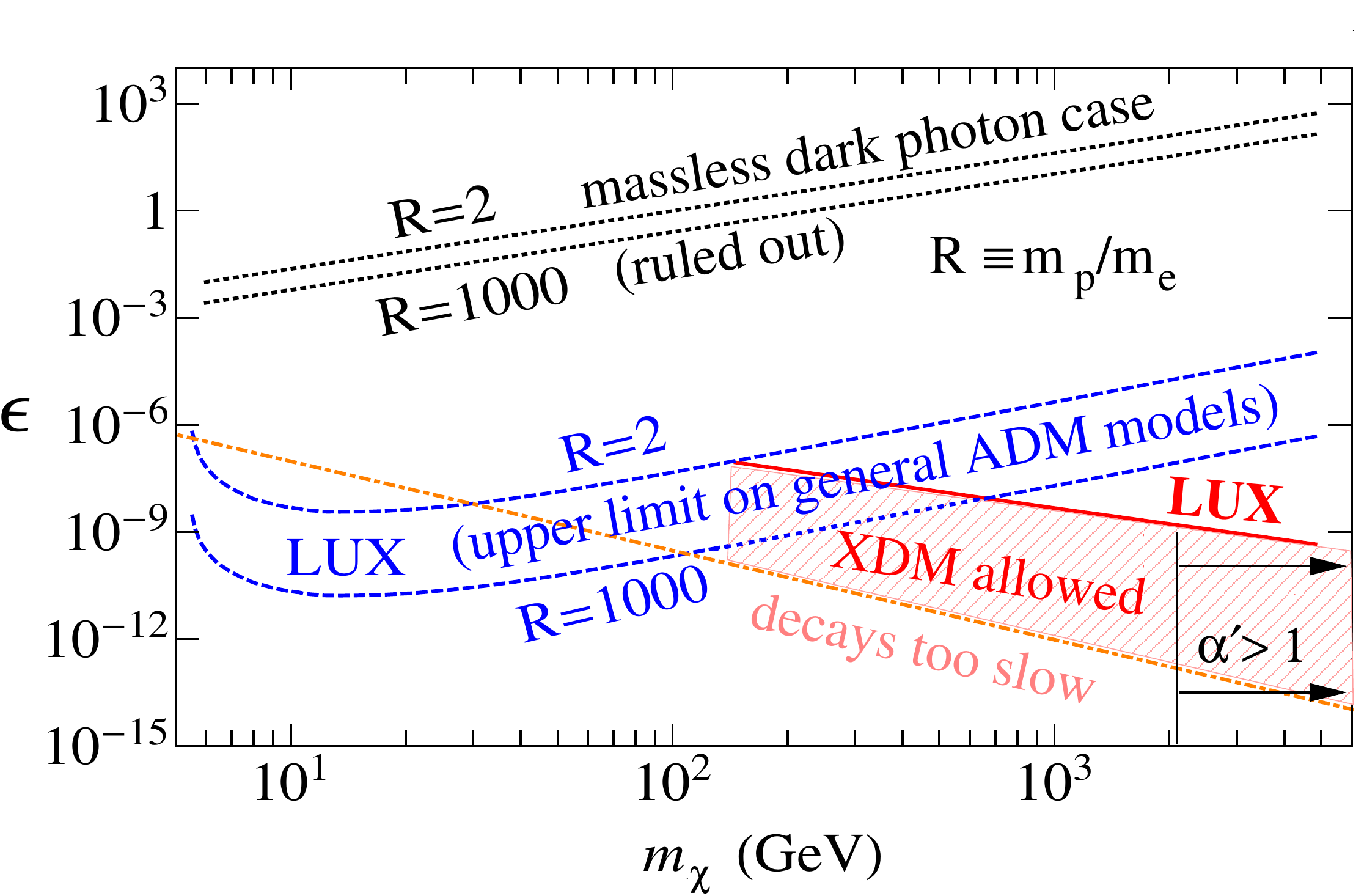}
\includegraphics[width=\columnwidth]{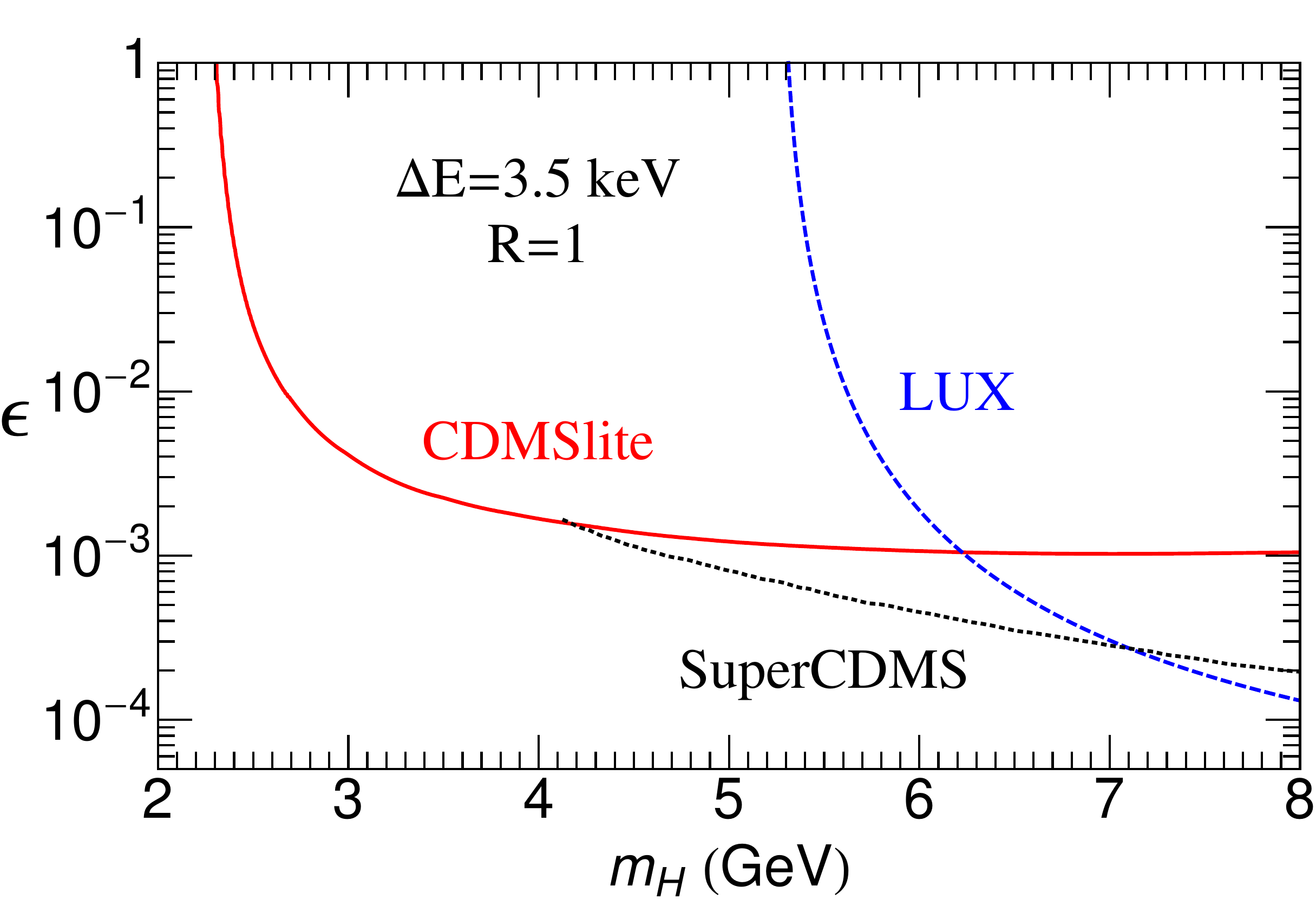}
}
\caption{Constraints on fractional charge $\epsilon$ of dark atom
constituents.  Left: lower (dashed) 
curves are LUX \cite{Akerib:2013tjd} constraint on $\epsilon$ for $R=2$ and $R=1000$, 
while the upper (dotted) lines 
show the values of $\epsilon$ required by the observed 3.5 keV X-ray 
line strength in the massless dark photon case.  The solid
line labeled ``LUX'' is the LUX upper limit (\ref{eps_lim})
on  $\epsilon$ in the massive dark photon model with $R$ fixed by
eq.\ (\ref{mdHval}).   Also shown (dot-dashed) is the
lower limit (\ref{eps_eq2}) from requiring fast decays of the excited
state (``decays too slow''), and the region where $\alpha'$ 
becomes nonperturbative.  
Right: constraints on $\epsilon$
from direct detection, in the case $R=1$ (equal dark proton and
electron masses), for $m_\dH < 8$ GeV. For SuperCDMS \cite{Agnese:2014aze}, only the region where it
provides a stronger constraint than CDMSlite \cite{Agnese:2013jaa} is shown. }
\label{fig:lux}
\end{figure*}

\smallskip
{\bf\noindent 2.\ Strength of X-ray line from scatterings.}  Ref.\ \cite{Finkbeiner:2014sja} shows
that the observed 3.5 keV X-ray line strength can be explained by 
mildly inelastic DM scatterings followed by fast decays of the excited state if
the cross section satisfies
\be
	\langle\sigma v\rangle\,BR \sim 10^{-21}{\rm\,cm}^3\,{\rm s}^{-1}
	\left(m_\dH\over{\rm\, GeV}\right)^2 % = {8\times 10^{-5}\,
	%m_\dH^2\over{\rm\, GeV}^{4}}
\label{FW}
\ee
(in units $\hbar=c=1$). Here $BR$ denotes the branching ratio for the
excited state to decay into visible photons as opposed to dark
photons.  If the dark photon was massless and mixed with 
the visible photon through the kinetic mixing term
$(\epsilon/2)F'_{\mu\nu}F^{\mu\nu}$, then the dark electron and proton
would get fractional charges $\mp \epsilon e$ and we would find that
$BR = \epsilon^2 \alpha/\alpha'$.  This scenario is ostensibly ruled 
out as we will explain in sect.\ 6.  Therefore we 
instead assume that the dark photon has a mass $m_{\gamma'} > 3.5$ keV
so that the hyperfine excited state can decay only into photons.
Hence we take $BR=1$.  

In sect.\ 9 we will show that the cross section for
spin excitations is of the same order as the elastic cross section,
which was found in ref.\ \cite{Cline:2013pca} to be of order
\be
	\sigma_{\rm el} \cong 100\, a_0^2 \equiv {100\over
	(\alpha'\mu_\dH)^2}\equiv {100\,f(R)^2\over 
	(\alpha'm_\dH)^2}
\label{sig_el}
\ee
where $a_0$ is the dark sector Bohr radius, 
$\mu_\dH = m_\pd m_\de/(m_\pd + m_\de)$ is the reduced mass, 
$R=  m_\pd/m_\de\geq 1$ and $f(R) = 
m_\dH/\mu_\dH= R + 2 + R^{-1}$.
It was shown that the coefficient estimated here as 100 
varies only moderately with $R$ for $R\lesssim 2000$.  With these definitions,
eq.\ (\ref{dE}) implies the constraint
\be
	\alpha' = 0.034\, (m_\dH/{\rm GeV})^{-1/4}\,f(R)^{1/2}
\label{alpha_eq}
\ee

Equating (\ref{sig_el}) to the cross section in (\ref{FW}), assuming a
relative velocity $v\sim 2000\,$km/s appropriate for galaxy clusters,
and using (\ref{alpha_eq}) to eliminate $\alpha'$
allows us to solve for the dark atom mass as a function of $R$:
\be
	m_\dH = 137\,[f(R)/4]^{2/7} {\rm\, GeV}
\label{mdHval}
\ee
We will see in sect.\ 8 that $R\gtrsim 100$ is needed to get efficient
recombination in the dark sector to form atoms. Then 
eq.\ (\ref{alpha_eq}) gives $\alpha'\gtrsim 0.08$.  Eqs.\
(\ref{alpha_eq}) and (\ref{mdHval}) together imply that $\alpha'>1$
for $R>5\times 10^4$, invalidating a perturbative treatment;
hence our preference for $R\lesssim 10^4$.

\smallskip
{\noindent\bf 3.\ Line strength from slow decays.}
If the lifetime $\tau_{hf}$ of the hyperfine excited state is greater than the
age of the universe $\tau_u$, the line strength is determined differently.
The dark atoms form with a 3:1 ratio of spin states
(since the temperature when they recombine is large compared to
$\Delta E$), and this ratio is preserved for of order 1 decay
time. So if $\tau_{hf}\ge\tau_u$, then a fraction of order unity
of the atoms is hyperfine-excited, {\it independently}
of the scattering rate.  Previous analyses showed that under these
circumstances, the observed line strength requires a decay rate of
\cite{Boyarsky:2014jta,Bulbul:2014sua}
%\be
	$\tau_d^{-1} = 2.3\times 10^{-21}{\rm\,s}^{-1}
	\left(m_\dH/100{\rm\,GeV}\right)$
%\ee
whereas the predicted rate is
\be
	\tau_{hf}^{-1} = {\alpha\epsilon^2\over 3}\, {\Delta
	E^3\over\mu_\dH^2}
\label{gamma_hf}
\ee
(Recall that we have assumed that the decays to dark photons
$\gamma'$ are kinematically blocked since $m_{\gamma'}> \Delta E$.)
Equating the two rates  gives the required value of the dark fractional
charge:
\be
	\epsilon = 1.2\times 10^{-14}\, (m_\dH/{\rm GeV})^{3/2}\, f(R)^{-1}
\label{eps_eq1}
\ee

\smallskip
{\bf\noindent 4.\ Constraints on fractional charges.} 
If the hyperfine state is longer lived than the age of the universe,
the required fractional charge (\ref{eps_eq1}) of $\de$ and $\pd$
is well below laboratory bounds.  Otherwise, we rely upon scatterings
to populate the excited state, and the requirement that $\tau_d <
\tau_u$ puts a lower bound on $\epsilon$,
\be
	\epsilon >  3\times 10^{-10}\, (m_\dH/{\rm 100\ GeV})^{-5/2}
\label{eps_eq2}
\ee
where we used (\ref{mdHval}) to eliminate $f(R)$.

There are upper limits on $\epsilon$ from the interactions of
dark atoms with nuclei by photon exchange.
(Due to interference between
visible and dark photon exchange, the parameter actually constrained
by direct detection is $\epsilon_{\rm eff} \cong \epsilon +
g'\delta/e$; see sect.\ 7 for definition of $\delta$.
Since $\delta$ is much less than the experimentally allowed value of
$\epsilon_{\rm eff}$, our conclusions are not changed by this
distinction, which we henceforth ignore.)  In ref.\  \cite{Cline:2013zca}
the value of $R$ corresponding to 
a given $m_\dH$ was determined by demanding strong 
self-interactions $\sigma/m_\dH \sim 1$b/GeV, more about which 
in sect.\ 5.
Here we 
recompute these  direct detection constraints as a function
of $m_\dH$, instead
using the relation (\ref{mdHval}) to determine $R$.  The resulting
limits are shown in
the left-hand graph of 
fig.\ \ref{fig:lux}.  
For dark atoms in the favored parameter range $m_\dH\sim 350-1300$ GeV and 
$R\sim 100-10^4$, the limit is $\epsilon 
\lesssim 10^{-8}$.  More precisely, 
\be
	\epsilon < 1.5\times 10^{-7}\left(100{\rm\ GeV}\over
	m_\dH\right)^{3/2} = 1.7\times 10^{-7}\, R^{-3/7}
\label{eps_lim}
\ee
where in the last equality we used (\ref{mdHval}) to trade
$m_\dH$ for $R$.  This is far above the lower bound
(\ref{eps_eq2}), as shown in fig.\ \ref{fig:lux}.

\smallskip
{\bf\noindent 5.\ Constraints on DM self-interactions.}  
There are numerous constraints from structure formation indicating
that self-interactions of cold dark matter should not have a cross
section exceeding $\sim 1$b per GeV of DM mass, whereas saturating
this limit may have beneficial effects for structure formation (see \cite{Weinberg:2013aya} for a recent
review).  Applying this bound to the cross section (\ref{sig_el}), using
eq.\ (\ref{alpha_eq}) to eliminate $\alpha'$, gives a lower 
bound on $m_\dH$:
\be
	m_\dH > 7.2\,(f/4)^{2/5}{\rm\ GeV} 
\label{SIDM}
\ee
which is satisfied by (\ref{mdHval}) for $R < 10^{12}$.   Thus our
model with fast decays is far from 
saturating the bounds on DM self-interactions.  
For representative values of $m_H\sim 350-700$ GeV and $R=100-1000$,
we find that $\sigma_{\rm el}/m_\dH = 1.5-3$ mb/GeV,
much less than the maximum allowed value
of $\sim 1$ b/GeV.   However the model with slow decays has
considerably greater freedom, making it possible to saturate
(\ref{SIDM}) there while still explaining the 3.5 keV line.

\smallskip
{\bf\noindent 6.\ Problems with massless dark photon.}
As was pointed out in ref.\ \cite{Frandsen:2014lfa}, if the 
dark photon is massless, the branching ratio $BR$ in eq.\ (\ref{FW}) is 
$\epsilon^2 \alpha/\alpha'$ rather than unity, and the 3.5 keV line
strength requires relatively large values of the kinetic mixing,
$\epsilon = 8\times 10^{-4} f^{-1/4}(m_\dH/{\rm GeV})^{13/8}$.
%\label{eps_eq}
The bound (\ref{SIDM}) from DM self-interactions requires $m_\dH> 7$
GeV, hence $\epsilon = 0.013$ for the $R=1$ case.
Fig.\ \ref{fig:lux}(right) shows that the CDMSlite results
\cite{Agnese:2013jaa}  imply  that 
such a large value of $\epsilon$ would require $m_\dH < 2.2$ GeV,
leading to $\sigma/m_\dH \sim 20$ b/GeV, twenty times the maximum
allowed value.  For $R>1$ the direct detection constraints become
stronger and the tension is exacerbated.  Using the
cross section for scattering on protons, $\sigma_p = 4\pi
(\epsilon \alpha m_p  a_0^2)^2$ \cite{Cline:2012is}, we obtain the
LUX limit \cite{Akerib:2013tjd} on $\epsilon$ shown in fig.\ 
\ref{fig:lux}(left) for the
range $2 < R < 10^3$.  There one sees that the limit is exceeded
by several orders of magnitude by the value  needed
to match the X-ray line strength.  For these
reason we consider the dark photon to be massive, with $m_{\gamma'} > 
3.5$ keV.
 
\bigskip
{\bf\noindent 7.\ Constraints on the dark photon.}   As long as the Compton
wavelength of the dark photon is greater than the required range
of its interaction, $m_{\gamma'} \ll a_0^{-1} = \alpha'm_\dH/f$, the photon mass will not 
significantly change the dark atom binding properties.  For our preferred
parameter values, this implies $m_{\gamma'} \ll 100$ MeV.  However,
for recombination we need $m_{\gamma'}$ to be less than the binding
energy so that excited states of the $\dH$ atom can radiatively decay
to the ground state.  This requires $m_{\gamma'} \ll \frac12\alpha'^2
m_\de \cong 7$ MeV (this numerical value applies in the models with
fast decays of the excited state). 

A simpler alternative to spontaneous breaking of
the gauge symmetry through a dark Higgs VEV is to suppose that 
the mass arises from the Stueckelberg mechanism, as
discussed in ref.\ \cite{Feldman:2007wj}.  This alternative also has string
theoretic motivations \cite{Ghilencea:2002da}.
The couplings of  the
visible and dark photons to matter in the two sectors is a function of
the kinetic mixing parameter $\tilde\delta$ and the ratio $\tilde\epsilon$ 
of Stueckelberg masses.  In the basis where the kinetic and mass
matrices are diagonal, the interactions become \cite{Feldman:2007wj}
\be
	A'_\mu(J'^\mu +(\tilde\epsilon-\tilde\delta)J^\mu) + 
	A_\mu (J^\mu -\tilde\epsilon J'_\mu)
\ee
where $A'$ ($A$) denotes the dark (visible) photon vector potential 
and $J'$ ($J$) the current of the dark (visible)
constituents.  (Couplings of the $Z$ boson to $J'$ are further
suppressed by the small ratio $(m_{\gamma'}/m_Z)^2$.)
Defining $\tilde\epsilon = (e/g)\epsilon$, the dark consituents 
get fractional charges given by $\pm\epsilon e$, while the dark gauge boson
couples to electrons with strength 
$(\tilde\epsilon -\tilde\delta)e\equiv \delta e$.  It would require
fine tuning $\tilde\epsilon =\tilde\delta$ to eliminate the coupling
$\delta$ 
between $A'$ and electrons, whereas $|\delta|$ could naturally be much 
larger than $\epsilon$ if $\tilde\delta\gg \tilde\epsilon$.  

However, various constraints on $\delta$ from astrophysics and
from fixed-target experiments 
are stronger than those on $\epsilon_{\rm eff}$
from direct detection.  We will be interested in $m_{\gamma'}\sim$ 
1 MeV
(which is still consistent with the need for excited dark atoms to 
radiate),
since constraints on $\delta$ are even more stringent at
lower $m_{\gamma'}$.  For $m_{\gamma'}\sim$ 1 MeV,  supernova cooling
restricts $\delta\lesssim 6\times 10^{-11}$ \cite{Dreiner:2013mua}.  
Then the theoretically most natural situation is that this constraint
is nearly saturated.  

One may wonder whether the dark photon can have an effect on big bang
nucleosynthesis (BBN) by its contribution to the energy density. If
$m_{\gamma'}>2 m_e$,  it decays into $e^+ e^-$ with rate $\Gamma_{ee}
= (1/3)\alpha\, \delta^2\, m_{\gamma'} g(x)$ where
$x=(m_e/m_{\gamma'})^2$ and  $g(x) = (1+2x)(1-4x)^{1/2}\sim 1$.  Then
for example with $m_{\gamma'}=3m_e$ and $\delta$ saturating the
supernova constraint, the lifetime is 54\,s which could be problematic
for BBN.  However we find that Thomson scattering $\gamma
e\leftrightarrow\gamma' e$ is too slow to ever bring $\gamma'$ into
equilibrium, given the constraint $\delta< 6\times 10^{-11}$.  The
corresponding process on dark electrons, $\gamma
\de\leftrightarrow\gamma' \de$, also fails to come into equilibrium
because of the limit 
 (\ref{eps_lim}) on $\epsilon$.  Thus the abundance of $\gamma'$ is
suppressed when it decays.  In sect.\ 10, we will estimate this abundance
based on assumptions about reheating of the dark sector, and show how
decays of $\gamma'$ into sterile neutrinos can be consistent with
hints from the CMB of an extra neutrino species, while simultaneously
satisfying BBN constraints.

\smallskip
{\noindent\bf 8.\ Dark recombination.}  Another important requirement for
the consistency of our proposal is that the dark constituents
recombined efficiently into atoms, with a sufficiently small residual ionized
fraction at the end.  The recombination of dark atoms has been
considered in references \cite{Kaplan:2009de} 
and \cite{CyrRacine:2012fz}.  Ref.\ \cite{Cline:2012is} numerically
fit the results of \cite{Kaplan:2009de} for the ionization fraction as
\vskip-0.6cm
\be
	X_{\de} \sim \left(1 + 10^{10}\alpha'^{4}\,{
	{\rm GeV}^2\over m_\de\, m_\pd}\right)^{-1}
\label{fi_eq}
\ee
\vskip-0.1cm\noindent
while \cite{CyrRacine:2012fz} obtained a similar estimate but with
an extra factor $\xi^{-1}$ multiplying $\alpha'^4$, where $\xi$ is the ratio of dark to visible photon 
temperatures, which suppresses
$X_{\de}$.  In sect.\ 10, we will estimate that $\xi=0.6$ in our model.
To be conservative, we use (\ref{fi_eq}) without this correction.
Eliminating $\alpha'$ and $m_\dH$ using (\ref{alpha_eq}) and 
(\ref{mdHval}), we find that $X_{\de}\sim 60\, R^{-15/7}$, so that
$X_{\de} < 3\times 10^{-3}$ for $R> 100$.  This is small enough to
consider that the dark constituents are mostly recombined, and the
unscreened dark Coulomb scattering of the ionized fraction will have a
negligible effect on the shape of dark matter halos as they evolve.

\smallskip
{\noindent\bf 9.\ Spin excitation cross section.} The cross sections for
hyperfine excitations are  related to those of elastic scattering
through \cite{zygelman}\\
\vskip-0.8cm
\be
	\sigma^\pm =  {\pi\over 2k^2}\sum_{\ell {\rm\ even}
	\atop \ell {\rm\ odd}}
	(2\ell + 1)\sin^2(\delta_t-\delta_s)
\ee
\vskip-0.2cm
where $\delta_{s,t}$ are the electron singlet and triplet channel
phase shifts that were computed for dark atoms in ref.\ 
\cite{Cline:2013pca}, and $\pm$ refers to $\Delta F = 2,1$ changes in
the total hyperfine state $F$ of the two atoms.  A linear combination
$(\sigma^++2\sigma^-)/4$ is relevant for finding the rate of 
hyperfine-exciting transitions.  For our purposes, it is
sufficient that $\sigma^+$ (which dominates) is of the same order as the unpolarized 
elastic cross section, which at low energies is dominated by the
$\ell=0$ phase shifts.

\smallskip
{\noindent\bf 10.\ Link to sterile neutrinos.}  It is possible
that the dark photon $\gamma'$ has an invisible decay channel in addition to
its $\delta$-suppressed decays into $e^+e^-$.  These decays could
therefore provide a source of dark radiation as well as avoiding
BBN constraints on $\gamma'$. 
The recent BICEP2 measurement of $B$-mode polarization in the CMB
\cite{Ade:2014xna}
reveals a tension with Planck determinations of the power spectrum
(for caveats, see \cite{Audren:2014cea}),
that can be alleviated if there is an additional radiation species
\cite{Giusarma:2014zza}.
In particular, sterile neutrinos with a mass $\sim 0.5$ eV have been
shown to give a good fit to the data 
\cite{Zhang:2014dxk,Dvorkin:2014lea,Archidiacono:2014apa}
(however ref.\ \cite{Leistedt:2014sia} disagrees with this conclusion). 
Here we consider the effect of
such particles coupling to the dark photon and whether they can
naturally contribute $\Delta N_{\rm eff}\sim 1$ to the effective
number of neutrino species.

Decays of $\gamma'$ into $\nu_s$ could arise from a one-loop
effect if there is a dark analog of the weak interactions coupling $\nu_s$
to $\de$ and a $W'$ boson with strength $g'_w$.  If $\nu_s$ is a Majorana particle, then
the effective operator induced is the anapole moment
\be
	\sfrac12 a_\nu\,\bar\nu_s\gamma^\mu\gamma_5\nu_s\,
	\partial^\alpha F'_{\alpha\mu}
\label{anueq}
\ee
where $a_\nu\cong{{g'_w}^{\!\!2} g'/( 16\pi^2
m_{W'}^2)}$.   For massive on-shell photons the 
Proca equation gives $\partial^\alpha
F'_{\alpha\mu}=m_{\gamma'}^2\, A'_\mu$, leading to the dimensionless coupling
$a_\nu m_{\gamma'}^2$ and a decay rate for $\gamma'\to\nu_s\bar\nu_s$ of
$\Gamma_{\nu\bar\nu} = a_\nu^2 m_{\gamma'}^5/24\pi$.

To find the $\Delta N_{\rm eff}$ of $\nu_s$ following these decays, we need
to know the abundance of $\gamma'$ when it decays: $Y_{\gamma'}=n_{\gamma'}/s$
where $s$ is the visible sector entropy density.  Previously we established that
$\delta$ and $\epsilon$ are too small for $\gamma'$ to come into
equilbrium with the standard model; hence we content ourselves with
making a reasonable
assumption about its initial temperature after reheating.  Namely if
the dark and visible sectors were initially heated to the same
temperature $T\gg 100$ GeV, and the dark sector consists only of $\de$, $\pd$,
$\gamma'$, $\nu_s$, $W'_a$ and a dark Higgs, then $Y_{\gamma'}=
(45\zeta(3)/2\pi^4)(13.74/106.75) = 0.036$, corresponding to a
temperature of $T_{\gamma'}/T_\gamma = (7\,Y_{\gamma'})^{1/3}= 0.63$. Here $13.74$ is the
number of dark entropy degrees of freedom that end up as $\gamma'$ as opposed
to $\nu_s$ (before $\gamma'$ decays), while 106.75 is the number of SM degrees of freedom.

Ref.\ \cite{Menestrina:2011mz} (see their fig.\ 3) has determined the contribution of
such a decaying dark photon to $\Delta N_{\rm eff}$ in the CMB as a
function of $Y_{\gamma}m_{\gamma'}$ and the lifetime of $\gamma'$
(assuming it decays only into dark radiation).  For example with
$Y_{\gamma'}=0.036$ and $m_{\gamma'}=1$ MeV, $\Delta N_{\rm eff}=1$
in the CMB is achieved if $\tau_{\gamma'}=40$\,s, while the
contribution to $N_{\rm eff}$ relevant for BBN is within the
constraints.  Such a lifetime is compatible with reasonable values of
the parameters entering (\ref{anueq}):  if $g'=1$ and
${g'_w}^{\!\!2}/4\pi=0.1$, then $m_{W'} = 17.5$\, GeV.  In this example
$\gamma'$ is too light to decay into $e^+e^-$ and
$\gamma'\to\nu_s\bar\nu_s$ is the only decay channel.  If $m_{\gamma'}=3
m_e$ with other parameters being the same, then $\delta$ need only 
be $<10^{-10}$ to allow decays into
$\nu_s$ to dominate over those into $e^+e^-$.

\smallskip
{\noindent\bf 11.\ Conclusions.}  We have outlined a scenario in which the
tentative discovery of a 3.5 keV X-ray line in galactic clusters
could be identified as the dark analog of 21 cm emission.  The dark
constituents have a small coupling $\epsilon e$ to ordinary photons. 
If the hyperfine excited state lifetime is greater than the age of 
the universe, very small values of $\epsilon$, eq.\ (\ref{eps_eq2}), are sufficient to 
give the observed line intensity, and the dark atoms could be
relatively light, $30-160$ GeV, having
strong self-interactions (eq.\ (\ref{SIDM})), which could alleviate problems of
cold dark matter with respect to structure formation at small scales.

Otherwise, for faster-decaying excited states, 
the observed X-ray energy and intensity, combined with the need 
to have a small ionized
fraction of the dark constituents and to satisfy direct dark matter
searches, more strongly constrain the parameter space of the model:
the dark atom mass is expected to be in the range $350-1300$ GeV, with
dark electron masses $3.5-0.1$ GeV, corresponding to dark 
proton-to-electron mass ratios $R\sim 100-10^4$, and U(1)$'$ couplings $\alpha'=
0.08-0.57$.  The fractional charges obey $\epsilon\lesssim 10^{-8}$,
which if saturated would make the dark atoms close to being discovered
by the LUX experiment.

Moreover the dark photon should have mass $m_{\gamma'}\sim$ 1
MeV to block the invisible decays of the hyperfine excited states and
allow them to decay only to visible photons  despite the smallness of
$\epsilon$.  The Stueckelberg mechanism is assumed to give rise to the
mass, as well as to a small coupling $\delta e$ of $\gamma'$ to
visible electrons with $\delta<6\times 10^{-11}$ from supernova
constraints.  Smaller values of $m_{\gamma'}$ could be permitted, at
the expense of stronger constraints on $\delta$.  In the 
fast-decay scenario, this would  require greater fine
tuning with respect to the theoretical expectation that $\delta
\gtrsim \epsilon$, whereas in the slow-decay models that expectation
is more easily satisfied.

Recent suggestions of a sterile neutrino of mass $0.5$ eV to reconcile
BICEP and Planck determinations of the CMB power spectrum can
naturally be incorporated into our dark sector.  An anapole moment
coupling of $\nu_s$ to the dark photon could allow for decays of
$\gamma'$ predominantly into $\nu_s$ rather than electrons, leading to
the desired density of $\nu_s$, compatible with BBN constraints.  Such a coupling would suggest that
the dark sector has a broken SU(2)$'$ gauge symmetry in analogy to the
weak interactions, in addition to the U(1)$'$ that binds the dark atoms.

{\noindent\bf Acknowledgments.}  We thank F.-Y.\
Cyr-Racine, G.\ Holder, K.\ Sigurdson  and Y.\ Zhao for helpful correspondence
or discussions. Y.F.\ acknowledges partial support from the  European Union FP7  ITN INVISIBLES (Marie Curie
Actions, PITN-GA-2011-289442) and  NORDITA.

%\bibliography{spibliography.bib}
% \bibliographystyle{h-physrev}
% \bibliographystyle{plain}
\end{document}